\documentclass[]{aa}
\usepackage{natbib}
\usepackage{psfig}
\def\158{$\lambda 158\mu$}
\def\63{$\lambda 63\mu$}
\def\ETC{{\it etc.}}

\def\EBV{\mbox{E$_{B-V}$}}
\def\HH{\mbox{H$_2$}}
\def\HD{{\rm HD}}

\def\nH2{{\rm n}({\rm H}_2)}
\def\NH2{{\rm N}({\rm H}_2)}
\def\pccc{{\rm cm}^{-3}} \def\pcc{{\rm cm}^{-2}}
\def\ccc{{\rm cm}^{3}} \def\pcc{{\rm cm}^{-2}}

\def\Tstar#1 {\mbox{${\rm T}_{\rm #1}^*$}}
\def\Tsub#1 {\mbox{${\rm T}_{\rm #1}$}}
\def\TK  {\Tsub K }

\def\p{\mbox{$^+$}}

\def\hcop{\mbox{{HCO\p}}}

\def\h13cop{\mbox{{H$^{13}$CO\p}}}

\def\c3h2{\mbox{C$_3$H$_2$}}
 
 \def\R0{R$_0$}

\def\ddeg{{}^\circ\kern-.1em}

\def\kms{\mbox{km\,s$^{-1}$}}
\def\ps{\mbox{s$^{-1}$}}

\def\E#1{\,10^{#1}}
\def\P#1,{$\nH2\TK~=~#1\times~10^4~\pccc$~K}
\def\ec#1,#2,#3,{#1\,(#2)\E{#3}}

\def\zoph{$\zeta$ Oph}

\def\H3p{\mbox{H$_3^+$}}

\def\zH{\mbox {$\zeta_{\rm H}$}}
\def\Nw{\mbox {N$_{\rm w}$}}
\def\DLAS{damped Lyman-$\alpha$ systems}
\title{Gas-phase recombination, grain neutralization and
cosmic-ray ionization in diffuse gas}

\titlerunning{gas, small grains and cosmic rays}

\author{H. Liszt\inst{1}}
\institute{National Radio Astronomy Observatory,
           520 Edgemont Road,
           Charlottesville, VA,
           USA 22903-2475}

\begin{document}
\date{received \today}
\offprints{H. S. Liszt}
\mail{hliszt@nrao.edu}

\abstract{
Atomic ions are mostly neutralized by small grains (or PAH molecules) 
in current theories of heating and cooling in cool diffuse clouds;
in the main they do not recombine with free electrons.  This alters 
the ionization balance by depressing n(H\p) and n(He\p) while carbon 
generally remains nearly fully once-ionized: charge exchange with atomic 
oxygen and formation of \HH\ and OH also depress n(H\p) in partly molecular 
gas.  Seemingly restrictive empirical limits on \zH\ are relaxed and 
higher values for \zH\ are favored in a wide range of circumstances, 
when grain neutralization is recognized.  Maintenance of the proton 
density at levels needed to reproduce observations of HD requires 
\zH\ $\ga 2\times 10^{-16}$\ps, but such models naturally explain 
the presence of both HD and \H3p\ in relatively tenuous H I clouds.
In dense gas, a higher ionization rate can account for high observed 
fractions of atomic hydrogen, and recognition of the effects of grain 
neutralization can resolve a major paradox in the formation of 
sulfur-bearing compounds.
\keywords{ISM: general; ISM: cosmic rays; ISM: atoms; ISM: molecules}
}
\maketitle

\section {Introduction.}

Several authors \citep{WolHol+95,Lis01,WeiDra01a,WelHob01} have commented recently 
on a seeming side-effect of current theories in which the heating of 
the diffuse ISM is dominated by the photoelectric effect on small grains
or PAH molecules \citep{Omo86,DHeLeg87,LepDal88a,BakTie94,WeiDra01Supp}.  The usual 
processes of direct radiative and  dielectronic recombination of atomic 
ions become secondary to neutralization during gas-grain interactions.  
When gas is somewhat denser or shielded from the interstellar diffuse soft 
X-ray flux, the proton density is much lower than would otherwise be the 
case and the total ionization fraction is much more nearly that given by 
the amount of free carbon in the gas alone; carbon remains mostly once 
ionized.   

This neutralization process was discussed by \cite{DraSut87}, \cite{LepDal88} and  
\cite{LepDal+88} and is not limited to diffuse gas or hydrogen ions.  Indeed, 
ion stage ratios like Mg II/Mg I, C II/C I \ETC\ should be entirely controlled 
in this way if large molecules {\it cum} small grains heat the neutral gas.  
All this notwithstanding, recognition of the importance of grain neutralization 
has occurred only with great hesitation, observational demonstrations of its 
relevance are still largely lacking and it is typically noted by observers 
(if at all) as something of a curiosity: \cite{WelHob01} is an obvious 
exception.  The original authors were concerned 
with establishing the existence of the effect and determining the abundance 
of the responsible agents, postulating what they described as {\it ad hoc} 
modifications of the chemistry.  Now, with the benefit of hindsight and with 
general acceptance of the heating effect of the same population of small 
grains, it is appropriate to recognize that the higher 
ion-neutralization rates associated with gas-grain interaction have other, 
largely unexplored consequences for our understanding of the interstellar 
medium (ISM).

The approach taken here to studying these effects is largely based on observations 
of various forms of hydrogen because they are uniquely sensitive to the cosmic-ray 
ionization rate in gas shielded from the diffuse soft X-ray flux.  Determination
of the proton density, cosmic-ray ionization rate and efficacy of grain 
neutralization are inextricably intertwined.   We show that some important 
historical limits on \zH\ (the cosmic-ray ionization rate of hydrogen, defined 
as the primary rate per neutral {\it atom}) must be restated, because they 
were derived only in terms of explicit ratios \zH/$\alpha_{\rm gas}$, where 
(symbolically) $\alpha_{\rm gas}$, the rate for gas-phase recombination with 
an electron, grossly underestimates the actual rate at which ions are 
neutralized.  Relaxation of these limits allows recognition that essentially
all observations of the various forms of hydrogen (protons, HD, \HH\ and 
\H3p\ in more diffuse gas; H I in dark molecular gas) suggest a much-enhanced 
cosmic-ray ionization rate.

The combination of faster neutralization of atomic ions -- which carry the
charge in diffuse gas -- and faster cosmic ray ionization has strong implications 
for the chemistry of diffuse, partly molecular gas.  Unlike atoms, molecular ions 
are dissociated by recombination more rapidly than they are neutralized by grains.  
If the cosmic-ray flux is enhanced, producing molecular ions more rapidly, without 
a concommitant increase in either the electron fraction or the molecular ion
destruction rate, the abundances of molecular ions may be much enhanced: HD and 
\H3p\ may coexist in the observed amounts, even in gas of fairly low density.

In the next Section, we lay out some of the underlying ionization and ion 
neutralization mechanisms in the context of recent two-phase heating/cooling 
models of the ISM.  In Sect. 3 we discuss some received limits on the proton 
density and cosmic ray ionization rate in atomic gas, and demonstrate
that limits on \zH\ from radio recombination lines are not restrictive if grain
neutralization occurs; we conclude that ionizing fluxes could be much 
higher than previously assumed.  In Sect. 4 we explore some consequences 
of the combination of grain neutralization and higher \zH\ for the chemistry 
of  diffuse clouds and we show that these include a much-enhanced abundance of 
\H3p, as has recently been seen in the ISM \citep{McCHin+02,GebMcC+99}, in the
same gas which harbors HD.  Section 5
is a brief discussion of processes in dark, fully molecular gas.

\section{Physical models of ionization, recombination and the ISM}

\subsection{Ionization in two-phase models of heating and cooling in the ISM}

Our discussion is based on a calculation of heating and cooling 
very similar to that set out by \cite{WolHol+95}; a spirited defense of 
classical notions of two-phase equilibrium, in the presence of other
influences, has recently been given by \cite{WolMcK+03}.  This
model has been used by us to discuss \HH\ and CO formation in diffuse
clouds \citep{LisLuc00}, the spin temperature of H I in warm gas
(it is not thermalized by collisions, see \cite{Lis01}) and the 
formation of \HH\ in \DLAS\ \citep{Lis02}.

Such models embody three primary sources of ionization and heating; soft 
X-rays, optical/uv photons and cosmic rays.  The X-ray flux is harsh but 
fragile and easily attenuated by relatively small gas columns.  For
the reference model a "bare" intrinsic spectrum is attenuated by a nominal 
gas column \Nw\ (following \cite{WolHol+95}; \Nw\ $\approx 10^{19}~\pcc$) 
whose value can be adjusted to reproduce the observed local mean density of 
electrons in intercloud gas (which we will also reference as pertaining to 
"free-space"), $\langle{\rm n_e}\rangle \approx 0.02~\pccc$ \citep{TayCor93}.  
It is a property
of two-phase models that the electron density in free space varies little 
with the total hydrogen density n(H) for 
n(H) $<$ $\langle{\rm n_e}\rangle / \xi_{\rm C}$ where  $\xi_{\rm C}$ 
is the abundance of free carbon in the gas phase.

The ionization rate of hydrogen required to ionize the intercloud gas 
at the observed levels is $10^{-15}$\ps, about 100 times 
larger than the presently accepted ionization rate due to cosmic rays.   
This intercloud ionization rate has changed little over many years
\citep{FieGol+69} and is probably the best-determined upper limit 
on acceptable \zH\ in unshielded regions. In cases with substantial
intervening gas columns, the X-ray flux is propagated accordingly in
our models; the spectrum hardens as the overall flux attenuates.

The photoionization rate can be parametrized in terms of a
scaling parameter G0, relative to a reference interstellar
radiation field, but this will not be varied here.  When gas 
parcels are modelled, the  attenuating column densities are
averaged appropriately over the assumed gas geometry.  For
\HH\ and CO, appropriate shielding factors are used \citep{LeeHer+96}.
In cool diffuse gas, the equilibrium temperature and pressure at a
given density are quite sensitive to G0, and to the assumed 
depletion of coolants (mainly carbon).

The reference value for the primary cosmic ray ionization rate of hydrogen
$atoms$ is here taken as $\zH = 10^{-17}$\ps.  The ionization rate
of a hydrogen $molecule$ is twice this and the total primary destruction 
rate of an \HH\ molecule due to cosmic rays, including dissociation, is 
$2\times1.08\times\zH$.  Table 1 gives the cosmic-ray ionization rates
of various species scaled to the reference value of \zH, taken from the 
UMIST database ({\it http://www.rate99.co.uk}), as are all the rate constants 
cited here, unless otherwise noted. 

%1
\begin{table}
\caption[]{Primary cosmic-ray ionization rates$^1$ and 
 undepleted relative abundances}
{
\begin{tabular}{lcc}
\hline
Species&$\zeta$ & [X/H] \\
  X    & ($10^{-17}$\ps) &  \\
\hline
H & 1.00 &  1.00 \\
D & 1.00 &  $1.5\times10^{-5}$ \\
He & 1.08 &  $7.9\times10^{-2}$\\ 
\HH & 2.00 &  \\
C & 3.83 & $3.6\times10^{-4}$\\
N & 4.50 & $1.0\times10^{-4}$\\
O & 5.67 & $7.5\times10^{-4}$\\
\hline
\end{tabular}}
\\
$^1$ from the UMIST reaction database
(http://www.rate99.co.uk) \\
\end{table}

\subsection{Grain neutralization}

The small grain population acquires electrons from the ambient gas by 
charge exchange with (neutralization of) positive ions having a superior 
ionization potential (typically, 6.8 eV according to \cite{Omo86}) and
loses electrons through the photoemission processes responsible for heating. 
In this way, the grain heating rate is dependent on the ionization level,
and both are controlled by the incident ionizing fluxes and particle density.

The neutralization rates follow from the grain charging formalism of 
\cite{DraSut87}, integrating over the grain size spectrum, assumed here, 
as in \cite{BakTie94} and \cite{WolHol+95} to be an extension of the 
classical MRN spectrum \citep{MatRum+77}; see the last paragraph in this 
subsection.  \cite{WeiDra01} give analytic
approximations to the low-temperature charging rates (not used here) and 
\citep{LepDal88} give a rate constant for cold gas.  Grain neutralization rate
constants for various ion species differ only inasmuch as their thermal 
velocities differ, {\it i.e.} all are proportional to the inverse square
root of the atomic weight.  

The abundance of small grains needed to heat the gas is about 100 times below 
the Solar abundance of carbon.  Thus, in diffuse gas where the ionization 
fraction is at least equal to the relative abundance of free carbon 
(typically 40\% of Solar) and grains lose electrons
rapidly to photoelectric emission, the charge is carried by atomic ions and free 
electrons.  In dense molecular gas where photoemission is nil and the equilibrium
gas-phase ionization fraction falls to values below $10^{-7}$, the negative charge 
on grains plays a substantial role in the overall charge balance \citep{LepDal88}.

\cite{WeiDra01Supp} have recently revisited the grain spectrum and its
effects on heating and cooling. They derive smaller sticking coefficients 
for a somewhat more copious small grain population and quote heating rates which 
are typically 20\%-50\% larger than those used here.  By contrast,
\cite{WolMcK+03} recently reduced the grain abundance and heating rate by a factor
two in their reference model for local gas. In light of this conflicting
guidance, we simply note that our assumptions appear to be intermediate.

\subsection{Charge exchange between oxygen and hydrogen}
%GolHab+69,FieGol+69,FieSte71,StaSch+99,ODoWat74,BarWal77

The most rapid gas-phase process affecting protons in atomic gas
is not recombination with electrons (or grain neutralization) but charge 
exchange with neutral oxygen.  As originally described by \cite{FieSte71} 
in the context of a purely atomic gas, the effect of this charge exchange 
was to change the O\p/O ratio from 10 n(H\p)/n(H I) to 
exp(-232 K/T) n(H\p)/n(H I), where the exponential represents the slight 
difference between the ionization potentials of H and O (the latter's 
being larger) and the higher ionization fraction of oxygen, 
absent charge exchange, follows from the 1971 version of Table 1.  
Basically, oxygen is expected to be ionized much more rapidly than 
hydrogen for a given cosmic-ray flux but is forced to donate much of 
the ionization to hydrogen.  The  effect on H was presumed to be 
relatively unimportant, owing to the smallness of the O/H ratio.  

Perhaps for this reason, O $+$ H charge exchange was not explicitly mentioned 
by \cite{WolHol+95} (it was included).  
However, the  displacement of charge  is accompanied 
by a slight diminution of n(p) in our models and inclusion of charge 
exchange becomes important when the proton density must be determined, 
especially if the abundance of molecular hydrogen is considered.  In even 
slightly denser gas having a molecular hydrogen component, O\p\ ions 
interact rapidly with $\HH$ to form O\H3p\ and this, like all molecular 
ions, recombines much more rapidly with electrons than does any once-ionized 
atom.  In the absence of charge exchange with hydrogen, the equilibrium 
abundance of O\p\ would be several orders of magnitude lower in gas 
having a  molecular hydrogen fraction of even a few percent.

%1
\begin{figure}
\psfig{figure=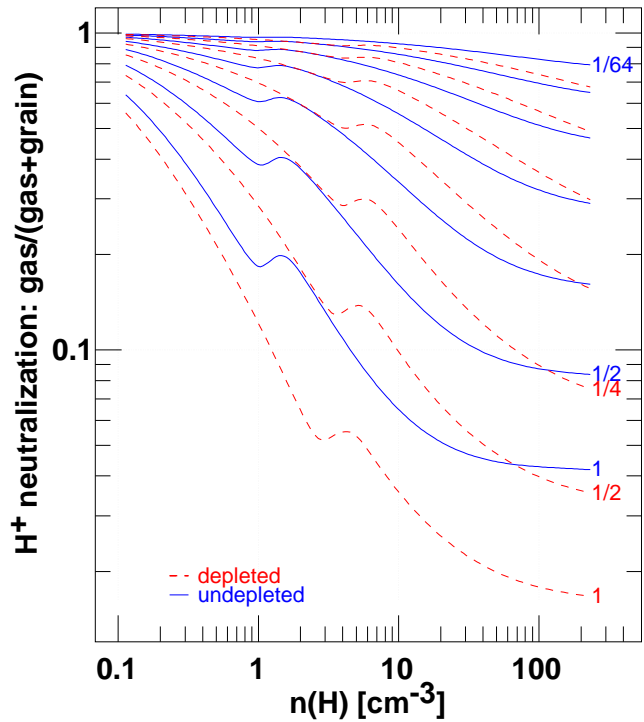,height=9.5cm}
\caption[]{The ratio of the effective gas-phase recombination rate 
n(e)$\alpha^{(2)}$ to the total rate at which  H\p\ ions are neutralized 
by gas-phase recombination and grain neutralization, in two-phase 
equilibrium.  The strength of the ion-grain interaction is artificially 
lowered in steps of 2 to 1/64th of the reference value.  Shown are results
for Solar and depleted ($\delta_{\rm C,N,O} = 0.4$) gas-phase abundances.}
\end{figure}

As noted by \cite{ODoWat74} and \cite{BarWal77}, maintenance of the O\p\ 
fraction can actually siphon charge from hydrogen when typical diffuse 
cloud chemistry operates more strongly in denser but still diffuse gas. 
This is another effect of molecules (the first being grain neutralization) 
which tends to lower the proton abundance in diffuse gas.  Although it is 
possible that a proton charge transferred to oxygen is returned to the gas 
after some chemical chains, this will not generally be the case. The 
upper limit on the neutralization rate of protons due to oxygen charge 
exchange is the rate at which O\p\ interacts with $\HH$ to form OH\p\ 
(the rate constant is $1.7 \times 10^{-9}\ccc$\ps) and our models simply 
assume that the process is either off (in so-called atomic gas) or on 
at this rate whenever the abundance of \HH\ has been calculated.  As 
shown below, converting 25\% of a moderate density cool gas to \HH\ 
results in a decline of 30\%-50\% in n(p).

\cite{StaSch+99} give the rate constant for the exothermic O $+$ H charge 
exchange channel as $3\times 10^{-10}\ccc$\ps below 100K, increasing as 
T$^{0.23}$ for T $>$ 100 K.  Increasing this rate has no strong direct 
effect on n(p) -- charge exchange already dominates -- but would increase 
n(O\p) somewhat in more strongly molecular regions, removing slightly 
more charge from hydrogen.

\subsection{Particulars and parameter sensitivities}

The reference model considered here assumes the small grain properties 
described by \cite{BakTie94};  we do not adjust the grain properties 
except artificially to demonstrate their specific influence on the gas.  
Figure 1 illustrates the efficacy of H-ion neutralization by grains, as 
used in the published calculations of two-phase heating and cooling 
{\cite{WolHol+95,Lis01,Lis02}. 
In this figure the per-grain cross-section for neutralization is 
artificially varied  away from the reference value (labelled 1); other 
aspects of the influence of small grains are not disturbed ($e.g.$, the 
heating rate).  The calculation is for free space, with and without gas-phase 
depletion of coolants (O,C) and charge sources (C).  Grain effects become 
stronger as the gas-phase ionization level declines (as could be the case 
in more heavily shielded regions).  Clearly, the usual gas-phase 
recombination processes play a distinctly secondary role in cooler gas 
and the total rate at which protons are neutralized is much higher than 
has typically been assumed previously.  
%Some care must be undertaken in
%interpreting this figure, as there are a variety of inter-dependencies
%(the models with depletion are also warmer) and the rates of various 
%processes have different dependences on the electron density.  

%2
\begin{figure}
\psfig{figure=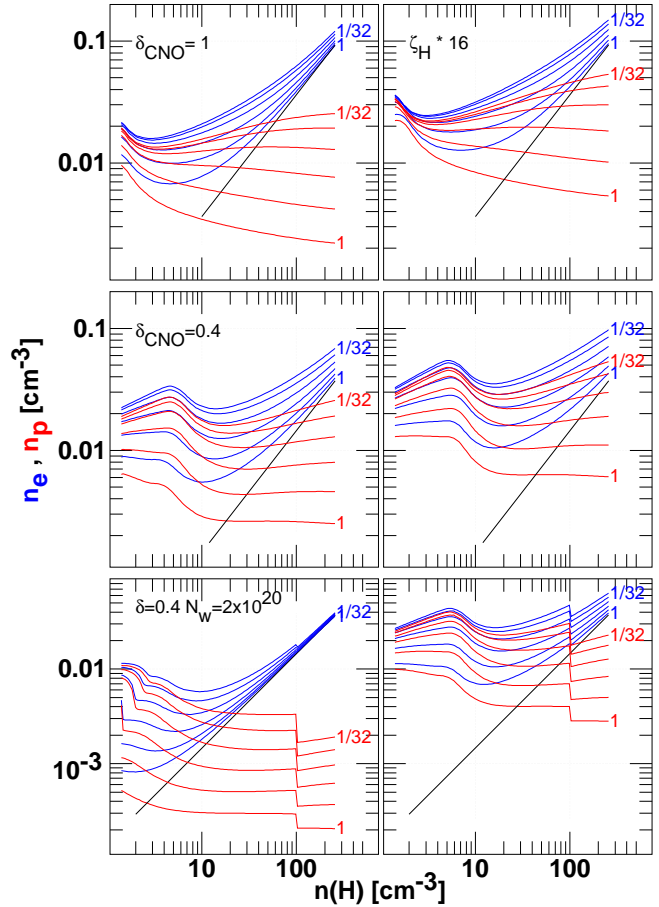,height=12cm}
\caption[]{Electron and proton density in two-phase
equilibrium for nominal (\zH $= 10^{-17}$\ps; left) 
and 16 times enhanced (right) cosmic-ray ionization rates 
under a variety of conditions. At top, for free space with 
Solar abundances; middle, free space with typical ISM 
depletions; bottom, with depletion, a substantial intervening 
neutral column density (\Nw) shielding the soft X-ray flux 
$and$, at high density (n(H) $>$ 100 $\pccc$),  
one-fourth of the H-nuclei in \HH. 
In each panel the strength of the grain neutralization interaction 
is artificially scaled from 1 times to 1/32 times the reference
value (steps of 1/2).  In each panel the proton density n(p) 
(shown red in appropriate media) is the family of curves 
which lies lower at right.}
\end{figure}

Shown in Fig. 2 are the electron and proton densities as functions 
of the total density, when the grain neutralization cross-section 
is altered  between 1 and 1/32 times the reference value, as in Fig. 1.
In the top panels, no elemental gas-phase depletion is included 
(corresponding most nearly to the phase diagrams of \cite{WolHol+95});
below, it is taken as a typical value.  In the right-hand panels, 
the rate of primary cosmic-ray ionization of atomic hydrogen is increased 
by a factor of 16 from its reference value of $1\times10^{-17}$\ps\ 
(all cosmic-ray rates scale in the same proportion; the rate of 
cosmic-ray ionization of O I, for example, is some 6 times that of 
H I, see Table 1).  The straight line in each panel represents 
n(C\p)$ = 3.6\times10^{-4}\delta_{\rm C}$ n(H) and departures of n(e)  
from it are due mostly to n(p) (He\p\ may contribute somewhat 
disproportionately at lower density, owing to its high cross section 
for ionization by X-rays).  In the lowest two panels, further 
complications are introduced; the gas is shielded from soft X-rays 
by half the hydrogen column density of a typical H I cloud 
\citep{Spi68,Spi78} and, for n(H) $> 100~\pccc$, a moderate 
fraction of the gas (one-fourth of the H-nuclei) is assumed 
to be molecular.

In cooler gas, neutralization of protons by small grains causes n(e) to 
decrease by factors of 2-3 under reference conditions,and n(p) decreases 
somewhat more.  The hydrogen emission measure product n(e)n(p) typically 
declines by a factor 10-20 in cool material, as suggested by Fig. 1.  The 
proton density follows \zH\ closely only when X-ray ionization is weak; 
otherwise, n(p) increases only very sluggishly with increasing \zH.

\section{Ionization balance in atomic gas}

\subsection{Limits on \zH\ in cool gas from radio recombination
lines}

Radio recombination lines (RRL) of hydrogen offer the possibility of 
fairly direct observation of the proton abundance.  To sense the cosmic ray
ionization rate, it is necessary to avoid lines of sight having any
contribution from H II regions and to select gas which is sufficiently
shielded that the X-ray ionization is weak.  The electron density
has a large contribution from n(p) at all densities in unshielded
regions if ion-grain neutralization is neglected, but only at low 
densities when it is included.  In practice it has turned out to be 
difficult to avoid gas having a large molecular fraction, although 
the complications introduced in this way are never taken into account
by observers.

Extant limits on the cosmic-ray ionization rate in cool atomic gas arise 
from (many) unsuccessful attempts to detect low-frequency (decametric) 
radio recombination lines (RRL) of atomic hydrogen 
\citep{Sha76,PaySal+84,PayAna+94}.  The basic analysis
begins with the condition of detailed balance for hydrogen

$$ n(e)n(p)\alpha^{(2)} = \Gamma_{\rm H} (n(H)-n(p)) $$

where $\Gamma_{\rm H}$ is the total ionization rate and $\alpha^{(2)}$ 
is the usual gas phase radiative recombination rate into the n $\ge 2$ 
levels, so that no ionizing photons are inadvertantly discounted.  The 
strength of a RRL is proportional to the line of sight integral  
$\int{n(e)n(p)}dl ~\equiv ~<n(e)n(p)>$L (L is some length), and to a generalized
function of the properties of the host medium and the atom, accounting
{\it inter alia} for non-LTE conditions among the level populations 
in the recombined atoms. 
The strength of the 21cm line is assumed to be proportional to 
$<$n(H I)$>$L = $<$(n(H)-n(p)$>$L over the same path 
and to a much simpler function of temperature (assuming that the
spin temperature is thermalized, a good assumption in cooler
gas).  In this case the equation of detailed balance is symbolically
recast as 

$$ {{\rm RRL~strength}\over{\rm 21cm~strength}} =
  {{\Gamma_{\rm H}}\over\alpha^{(2)}} F(n(H),n(e),T, \nu, etc.) $$.

The function F may have zeroes for certain combinations of density, 
temperature and quantum level, so the RRL to be studied must be chosen 
with some care.  For his model of the recombination, \cite{Sha76} derived 
limits $\zH \la 2-4 \times 10^{-16}$\ps\ in a dozen or so H I features seen
at low latitudes in the inner Galaxy.  Many of these are fairly distant 
(as judged by their velocities) and most are sources of CO emission in 
galactic surveys, so they must be fairly dense;  neglect of molecular
processes somewhat overstates the sensitivity of the RRL measurements to
\zH.  \cite{PaySal+84} set more stringent limits 
\zH\ $ < 0.4 - 1.6 \times 10^{-16}$\ps\ in gas toward 3C123 which is also
a well-studied molecular line source \citep{Cru80}. These limits have been
considered a significant demonstration that the cosmic ray ionization
rate is incapable of ionizing the intercloud gas to the observed
degree (see Sect. 2.1).
   
In the context of the current models for ion neutralization, such limits 
must be revised upward to the same extent that the effective recombination 
rate increases (Fig. 1), which in general makes them uninteresting.  Including 
grain neutralization reduces n(p) by factors of 6 or so at n(H) $\ga$ 
30 $\pccc$, and n(e) by factors of 2-3, which is sufficient to ensure 
that the sought-after hydrogen RRL could never be observed. The RRL data 
really cannot  be used to derive interesting limits on \zH, if it is accepted 
that grain neutralization is important.

Eventually it was recognized, largely on the basis of the behaviour 
of low-frequency carbon RRL (which $were$ seen), that the function
F(n(H),n(e),T, $\nu$, \ETC) is subject to considerable uncertainty
in terms of the model which is assumed for detailed balance in
the recombining atoms.  \cite{PayAna+94} analyzed low-frequency 
carbon RRL and upper limits on the optical depth of the hydrogen 
RRL toward Cas A, and did a quite detailed heating and cooling 
calculation of two-phase equilibrium (excluding grain neutralization 
processes).  Their limit on \zH\ for one feature with an inferred 
density n(H) = 200-500 $\pccc$, \zH\ $< 6 \times 10^{-17}$\ps, is 
perhaps less interesting now than the somewhat more robust inference 
that n(p)/n(e) $<$ 0.3.  In unshielded atomic gas, such low proton fractions 
cannot be achieved {\it unless} grain neutralization dominates.  However,
the gas seen toward Cas A is dense and dark enough that it cannot
be considered to be either diffuse, or atomic \citep{WilMau+93,LisLuc99}.  
The cosmic-ray ionization rate \zH\ is not meaningfully constrained by 
the Cas A data either, if grain neutralization is recognized.

\subsection{Determination of n(e) via optical spectroscopy}

%3
\begin{figure}
\psfig{figure=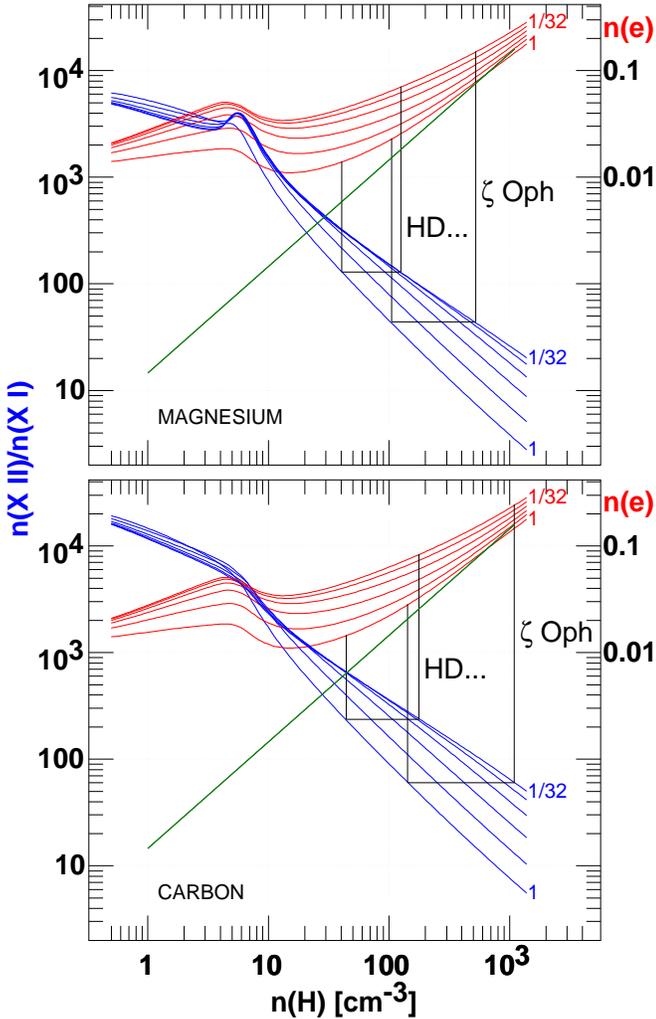,height=13.5cm}
\caption[]{Electron density (right scale; red in appropriate
media) and ion stage ratios n(X II)/n(X I) for X = magnesium (top) 
and carbon, with substantial gas-phase depletion of carbon and 
oxygen $\delta_{\rm C} = \delta_{\rm O} = 0.4$.  The ion stage
ratios (shown blue in appropriate media) are the families of 
curves which are lower at right; families are created by scaling 
the grain neutralization as in Figs. 1 and 2. The ion ratios observed 
toward \zoph\ and HD 192639 (see Sect. 3.2) are indicated by horizontal 
lines and the inferred hydrogen and electron densities are given 
by the vertical line segments connecting the families of curves
at these densities.}
\end{figure}

In their work introducing the notion of grain neutralization, 
\cite{LepDal+88} discussed its influence on commonly-occuring
measurements of ion ratios (Mg II/Mg I and the like); the basic idea was
that lowering such ratios with grain processes mimics the effect of higher
electron and total particle densities.  This in turn may help to 
reconcile discrepancies between densities measured from ion ratios and 
those gleaned from study of the excitation of C I fine-structure or
CO rotational levels.  This mechanism has, however, only been accepted
very grudgingly by observers of ion stage ratios, who are more likely
to pay lip service than to calculate the effect, and most likely to
sweep the matter under the rug.

Ratios of atoms in different stages of ionization depend on the total 
electron density, and would be sensitive to large n(e)-n(C\p) $\approx$ 
n(p);
this dependence is complicated and changed by grain neutralization, 
of course. If it were found that n(e) $>>$ n(C\p) in sufficiently dense 
gas (the ionization is supported by hydrogen at low density anyway), this 
would suggest that n(p)/n(C\p) was large, and that \zH\ was large enough 
to overcome whatever recombination/neutralization processes were occurring.  
In terms of Fig. 2, we see that this really cannot happen if grain 
neutralization occurs substantially as indicated, so high proton densities 
in sufficiently dense gas could be a strong contraindication of grain
neutralization.  
%Conversely, although grain neutralization can keep n(p)/n(e) 
%small in dense gas even when \zH\ is very large, n(p) $\ga$ n(C\p) 
%for n(H) $\ga 100~\pccc$ even when $\zH\ = 2 \times10^{-17}$ \ps. 
%So small n(p)/n(e) would require either strong grain neutralization 
%or $extremely$ small \zH.

Toward \zoph, \cite{SavCar+92} measured N(Mg II)/N(Mg I) = 105 and 
N(Fe II)/N(Fe I) = 537 for the denser gas at -15 \kms, and found 
n(e) $= 0.065~\pccc$ and $= 0.046~\pccc$ from the two ratios, 
respectively, based on a rather simple analysis of ionization 
equilibrium in an atomic gas in free space, neglecting grain 
neutralization and cosmic-ray ionization.  For n(p) = 0, these values 
of n(e) would be reproduced for n(H) = 300-400 $\pccc$, 
n(C\p)/n(H) = $1.5\times10^{-4}$.  These densities are very nominally 
appropriate to the feature in question, perhaps making it appear that 
n(p)/n(C\p) is not large.  However, a self-consistent application of the 
analysis in \cite{Spi78} shows that n(p)/n(C\p) would be of order unity 
even for \zH $= 10^{-17}$ \ps\ and n(H) a factor of two smaller, neglecting 
grain neutralization. 
In any case, \cite{SavCar+92} remarked that their values of n(e) 
were an order of magnitude smaller than in the classic analysis 
by \cite{Mor75}, which was based in part on the N(C II)/N(C I) 
ratio (now measured at 60 but previously ($ibid$) was 30).

The close agreement in n(e) from Mg and Fe ionization ratios
was illusory, as relevant f-values for the Mg II lines were
subsequently revised upward by a factor 2.4 \citep{Fit97},
lowering the derived Mg II/Mg I ion ratio and requiring a 
substantial upward scaling of the electron density derived from 
it.  Ion stage ratios seen toward \zoph\ are given in Fig. 3
here, which shows n(X II)/n(X I) for X = magnesium and carbon in 
a free-space, two-phase heating and cooling calculation
assuming a substantial depletion of gas-phase carbon and oxygen 
$\delta = 0.4$ so that [C]/[H] $= 1.44\times 10^{-4}$. 

In each panel of the figure the ion ratios are the families
of curves which lie lower at right and higher at left; the other 
family in each panel is the electron density.  Families of curves 
were created by scaling the grain neutralization rates (as in Figs. 
1 and 2) from 1 to 1/32 times the reference value.
The shaded trapezoidal outlines in each figure illustrate where 
the models reproduce measured ion ratios (the horizontal base)
and what electron densities are found with and (largely) without
grain neutralization.  The n(e) values cited by \cite{SavCar+92}
are reproduced nearly exactly for the ion ratios they quoted.

If grain neutralization is ignored, the hydrogen (electron) densities 
corresponding to the Mg and C ion ratios are  530 (0.15) and 1140 
(0.25) $\pccc$ respectively.  At the temperature inferred for this
gas from the ratio of population in the J=1 and J=0 levels of \HH, 
T = 56 K, the thermal pressure found from these ion ratios would be 
much higher (10-20 times) than inferred from excitation of the 
fine-structure levels of C I \citep{JenJur+83} or the rotational
excitation of CO \citep{Lis79,LisLuc94}.  Including grain 
neutralization greatly reduces the densities, to 105 (0.023) and 
145 (0.028) $\pccc$ , so that n(C\p) $\ga 0.66$ n(e) and the n(H)-T 
products derived from different analyses largely agree.

\cite{SonFri+02} present new ion stage ratios for several elements 
toward the low (mean) density but translucent (\EBV\ = 0.64 mag) line 
of sight to HD 192639.  Their ratios for carbon and magnesium are 
repeated in Fig. 3; in their analysis n(e) $\approx 0.1~\pccc$ for 
either species, slightly
higher than here.  Along this line of sight there is a large discrepancy
between the local density derived from excitation of the C I fine
structure levels, which indicates n(H) = 10-20 $\pccc$ for T = 100 K,
and the substantially higher densities n(H) $\approx 120-200~\pccc$ 
indicated by the ion stage ratios. \cite{SonFri+02} ascribed the 
seemingly high electron fraction to
a mixture of H I and H II gas along the line of sight but noted
that their analysis might be modified by grain neutralization.  Now 
we see that grain neuturalization does indeed remove the discrepancy.

\cite{WeiDra01a} assessed existing ion stage ratios in some clouds
of low inferred (from C I excitation) density n(H) = 10 - 20 $\pccc$ 
observed by  \cite{FitSpi97} and found that reductions in n(e) and the scatter 
among the n(e) indicated by differing ions could be achieved, but that 
discrepancies between electron and particle densities derived from
Ca III/Ca II and Mg II/Mg I persisted.
 
\cite{WelHob01} discussed a large survey of K I spectra and, in
the absence of K II spectra,  derived statistical properties of
the cloud ensemble based on a mean value for the depletion
$\delta_{\rm K}$.  The median electron density n(e) $= 0.1~\pccc$
is typical of the results discussed here.  These authors discussed
grain neutralization at some length and concluded it  was responsible 
for a factor three increase in N(K I), reconciling various pressure 
estimates.  \cite{WelHob01} noted Hobbs' long insistence that n(e) 
$\approx$ n(C\p) for clouds seen in Na I, K I, $etc$, which we know 
now is only possible in diffuse gas if grain neutralization dominates.

\section{Ionization balance in diffuse gas with appreciable \HH}

\subsection{Observations of HD}

HD does not partake of the same self-shielding which is the {\it sine 
qua non} of high $\HH$-fractions.  Hence the ratio n(HD)/n($\HH$) might 
be expected to be very small in diffuse gas, well below the inherent [D/H] 
ratio (locally, [D/H] $= 1.5\times10^{-5}$ as quoted by \cite{MooCow02}). 
That this is 
not so is due to fractionation and charge exchange processes which, however 
uncertainly, nonetheless allow an inference of n(p) without explicit
reference to the processes by which protons are neutralized 
\citep{BlaDal73,Wat73,Jur74,ODoWat74,Spi78}.

In a purely atomic gas ionized by cosmic rays the ionization and 
recombination rates of H and D atoms would be very nearly the same
(the grain neutralization of D\p\ is slower by a factor $\sqrt{2}$)
but a strong and slightly endothermic charge exchange with 
protons H\p $+$ D $+\Delta$E $\rightarrow$ D\p $+$ H
(rate constant $k_1 = 10^{-9}~\ccc$\ps, $\Delta$E/k = 41 K) 
tends to force n(D\p)/n(D I) $\approx$ n(H\p)/n(H I) exp(-41 K/T).  In the 
presence of $\HH$ a rapid and relatively strongly exothermic 
reaction D\p\ $ + \HH \rightarrow \HD\ + $H\p\ forms HD with rate
constant  $k_2 = 2.1 \times 10^{-9}~{\rm cm}^3$ \ps .
\footnote{rates in this Section are taken from Table 1 of \cite{StaLep+98}}
%so that n(D\p)/n(D) $\approx$ (n(p)/n({\rm H I})) exp(-41/T)/ 
%(1+k$_2{\rm n}(\HH)/{\rm k}_1{\rm n}({\rm H I})$)
%The volume formation rate of HD is k$_2$n(D\p)n(\HH) ($\pccc$ \ps) 
%and the destruction rate due to photodissociation is $\Gamma_{\rm HD}$ 
%(\ps) n(HD).  

If only charge transfer and \HH\ fractionation neutralize D\p\ a 
relatively compact expression gives n(p) in terms of observed 
quantities and physical constants, with no explicit dependence 
on either the density or recombination rates ({\it ibid}). We have 

$$ n(p) = {{ {{\rm N(HD)}}/{\rm N}(\HH)} \over {\rm [D/H]}}
  ~ { \Gamma_{\rm HD} \over{{\rm k}_2}} 
     [1+({{{\rm k}_2}\over {{\rm k}_1}}-2) {{{\rm n}(\HH)}\over{\rm n(H)}}] 
    \exp{ ({ {41} \over T}) } 
\eqno(1) $$

where n(H) = n(H I) + 2 n($\HH$) 
and the photodissociation rate of HD 
$\Gamma_{\rm HD} = \Gamma_{{\rm H}_2} = 5\times10^{-11}$\ps\ under 
typical conditions in free space \citep{LeeHer+96}. 
%older analyses 
%typically assumed a larger average photodissociation rate, 
%$\Gamma_{\rm HD} = 10^{-10}$\ps.  
The derived n(p) is nearly independent of the molecular fraction in 
the gas for k$_2/$k$_1 = 2.1$ and there is no explicit dependence
on density (only the \HH\ fraction).

%4
\begin{figure}
\psfig{figure=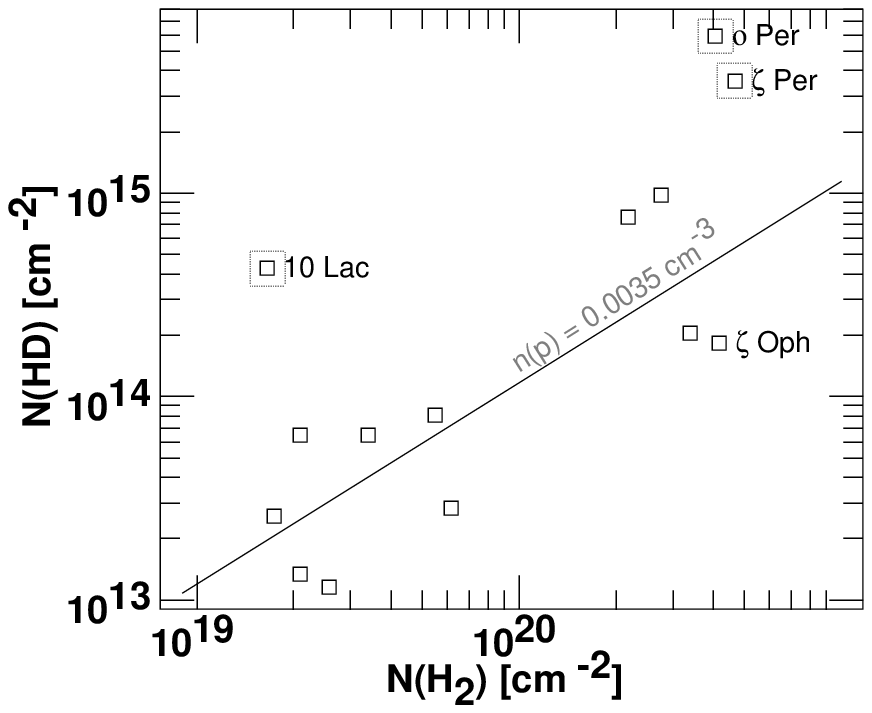,height=7.2cm}
\caption[]{HD and \HH\ column densities. N(HD) values are from 
\cite{SpiCoc+74} except for $o$ Per from \cite{Sno75}, $\zeta$ Per 
from \cite{Sno77} and \zoph\ from \cite{WriMor79}.  N(\HH) values
are from \cite{SavDra+77}.  The line labelled n(p) = 0.0035~$\pccc$
is from Eqn. 1, evaluated for T = 75 K and 
$\Gamma_{\rm HD} = 5\times 10^{-11}$\ps\ as in free space.}
\end{figure}

The available {\it (Copernicus)} observations of HD in the nearby ISM are 
summarized 
in Fig. 4; N(HD) values are from \cite{SpiCoc+74} except for $o$ Per from 
\cite{Sno75}, $\zeta$ Per from \cite{Sno77} and \zoph\ from \cite{WriMor79}.
The $\pm1\sigma$ error range for $\zeta$ Per is large, 0.5 dex.  With the obvious
exception of the three datapoints outlined (10 Lac, $o$ Per and $\zeta$ 
Per) the regression line fit to the data, log N(HD) $= -5.73 + 0.99$ log N(\HH),
has very nearly unit slope corresponding to uniform physical conditions
and N(HD)/N(\HH) $\approx 1.0-1.2\times10^{-6}$.  From Eqn. 1 we find 
n(p) $\approx 0.0035~\pccc$ at T = 75 K in free space.

Thus the nominal proton densities inferred from N(HD) are small compared to 
expected values of n(e) for all but the most discrepant datapoints,  and
may be reproduced in slightly darker gas lacking X-ray ionization {\it either}
with nominal cosmic ray ionization rates in the absence of grain neutralization,
{\it or}, with it, if \zH\ is substantially enhanced (Fig. 2 at bottom).
Earlier analyses attempted to infer \zH\ from n(p) and 
found an enormous range, for instance from $7 \times 10^{-18}$\ps\  for
$\xi$ Per to $10^{-15}$\ps\ for 10 Lac \citep{ODoWat74}.  

\subsection{Observations of \H3p}

The detection of \H3p\ in diffuse gas has had some ironic twists.  
Chemical models employed in the original discussion of grain neutralization
\citep{LepDal88,LepDal+88} were poisoned by inclusion of the then-fashionable 
assumption of a very small low-temperature gas-phase recombination rate for 
\H3p.  This led the authors to a gross overprediction of N(\H3p), which they 
suggested would be observable; their suggestion seems not to have been acted 
upon in a timely manner, thus depriving the world of a seeming corroboration 
of the incorrect recombination rate.  Instead, the 
recombination rate was corrected \citep{Ama88,LarDan+93,SunMow+94}, lowering 
expectations for the presence of \H3p , and, by the time it was widely 
detected, this was considered surprising.  With substantial fluctuations, 
N(\H3p)/\EBV $\approx 8\times10^{14}~\pcc/6$ mag (see
Fig. 14 of \cite{McCHin+02}), or N(\H3p)/N(H) $\approx 2.3\times 10^{-8}$
for N(H)/\EBV $= 5.8\times 10^{21}~\pcc$.  The clear increase in column
density with line of sight extinction strongly argues that \H3p\
is widely-distributed in a relatively common cloud component of the ISM,
not segregated in clouds of too-high density (which would be relatively rare).

The abundances of HD and \H3p\ really cannot be considered except in the 
broader context of a model for \HH\ itself (see the next subsection), but 
the  chemistry of \H3p\ is quite simple in broad outline.  Very nearly 
every ionization of an \HH\ molecule leads to formation of \H3p, which 
in diffuse gas is destroyed mainly by recombination with an electron. 
The rate constant for this process is taken as 
$\alpha_3 = 1.2\times10^{-7}~\ccc $(300/T)$^{0.5}$\ps, which is
our interpolation among experimental values which agree marginally at 
300 K and have somewhat different temperature dependences. 
\H3p\ is destroyed to a much smaller extent (typically, 100 times) 
by reaction with C and CO (rate constant $2\times10^{-9}~\ccc$\ps) 
or O (rate constant $0.8\times10^{-9}~\ccc$\ps).  The CO fraction in 
diffuse gas is quite small ($cf.$ Fig. 8 of \cite{LisLuc00}), and
grain neutralization at a rate 1/$\sqrt{3}$ that of H\p\ is also
negligible in diffuse gas.

The rate equation determining the \H3p\ abundance is typically cast 
as n(\H3p) n(e) $\alpha_3 \approx 2$\zH n(\HH) but the 
implicit dependences of n(\HH) and n(e) upon \zH\ prevent
n(\H3p) from increasing linearly with \zH\ for large \zH\
as suggested by Eqn. 3 of \cite{McCHin+02}.  In our model we
actually solved a small reaction network for the self-consistent
determination of the abundances of the hydrogen-bearing
species, including gas-phase formation of \HH\ via H$^-$.

\subsection{\H3p\ and HD in the context of \HH\ formation}

Figure 5 shows the run of \HH, HD and \H3p\ in simple diffuse
cloud models such as we have reported on in \cite{LisLuc00}
and \cite{Lis02}.  A spherical clot of gas is immersed in the same
conditions for which the calculations of two-phase equilibrium are
performed and molecular hydrogen in the uniformly illuminated gas 
is allowed to self-shield.  The \HH\ abundance is solved iteratively
in 64 radial zones along with the temperature, electron density,
 HD, and \H3p\ abundances \ETC.  Represented at top in the Fig. 
5 is something akin to a standard Spitzer H I cloud; at bottom is a model 
with four times higher number and column densities.  At right, the small 
grains have been treated differently at the same hydrogen number and 
column density; at top, their neutralization activity (alone) is 
suppressed by a factor 32.  At  bottom they are removed from the 
calculation to this degree.  The normalization is 
different for \H3p\ and HD because N($\HH$) is generally not known 
along lines of sight where N(\H3p) was detected (only N(H) is inferred, 
from the selective extinction using a standard gas-grain conversion).

Figure 5 shows that the observed abundances of HD and \H3p\ are 
both reproduced in the models at left, with grain neutralization
and a much-elevated (factor of 10-20) cosmic ray ionization 
rate.  In the absence of grain neutralization, {\it either} the typical 
HD/\HH\ ratio can be readily reproduced with the reference value of 
\zH, {\it or} \H3p\ can be reproduced with a much (20-100 times) 
higher value of \zH, at the cost of grossly overproducing HD.

Because the HD abundance is sensitive to the ionization level,
it is seen to increase toward the cloud edge (the effect is only
barely resolved); this would seem a
natural explanation for the high HD/\HH\ ratio toward the low
\HH\ column density line of sight toward 10 Lac.  It is also possible 
that HD is actually on the verge of becoming self-shielding.  Observed
values of N(HD) are already larger than those of N(\HH) at the knee
of the N(\HH) $vs.$ N(H) plot (for instance Fig. 8 of \cite{LisLuc00}). 

The higher ionization rates seemingly demanded by a distributed \H3p\
component have been noted by \cite{McCHin+02} and \cite{vanvan+00}. 
The latter authors also derived values of \zH\ from observations of 
H$^{13}$CO\p\ in the dense clouds associated with the same protostars 
used as background sources for the \H3p\ spectra, quoting 
\zH $= 2.6 \pm 1.8\times10^{-17}$\ps.  Noting the discrepancy in \zH,
they ascribed the increase of \H3p\ with \EBV\ to intervening clouds of 
density n(H) $\la 10^4~\pccc$, in which carbon was mainly in the form of
either C I or CO.  We would take strong exception to this view, which
does not account for HD and \H3p\ in the same gas, but the cosmic ray
ionization rate in dense gas was subsequently revised by \cite{Dotvan+02},
who quote \zH\ $= 5.6 \times10^{-17}$\ps with a factor three
uncertainty.  This does not exclude the higher values of \zH\ preferred
here.  The models used by these authors do not seem to have included
grain neutralization.

%5
\begin{figure*}
\psfig{figure=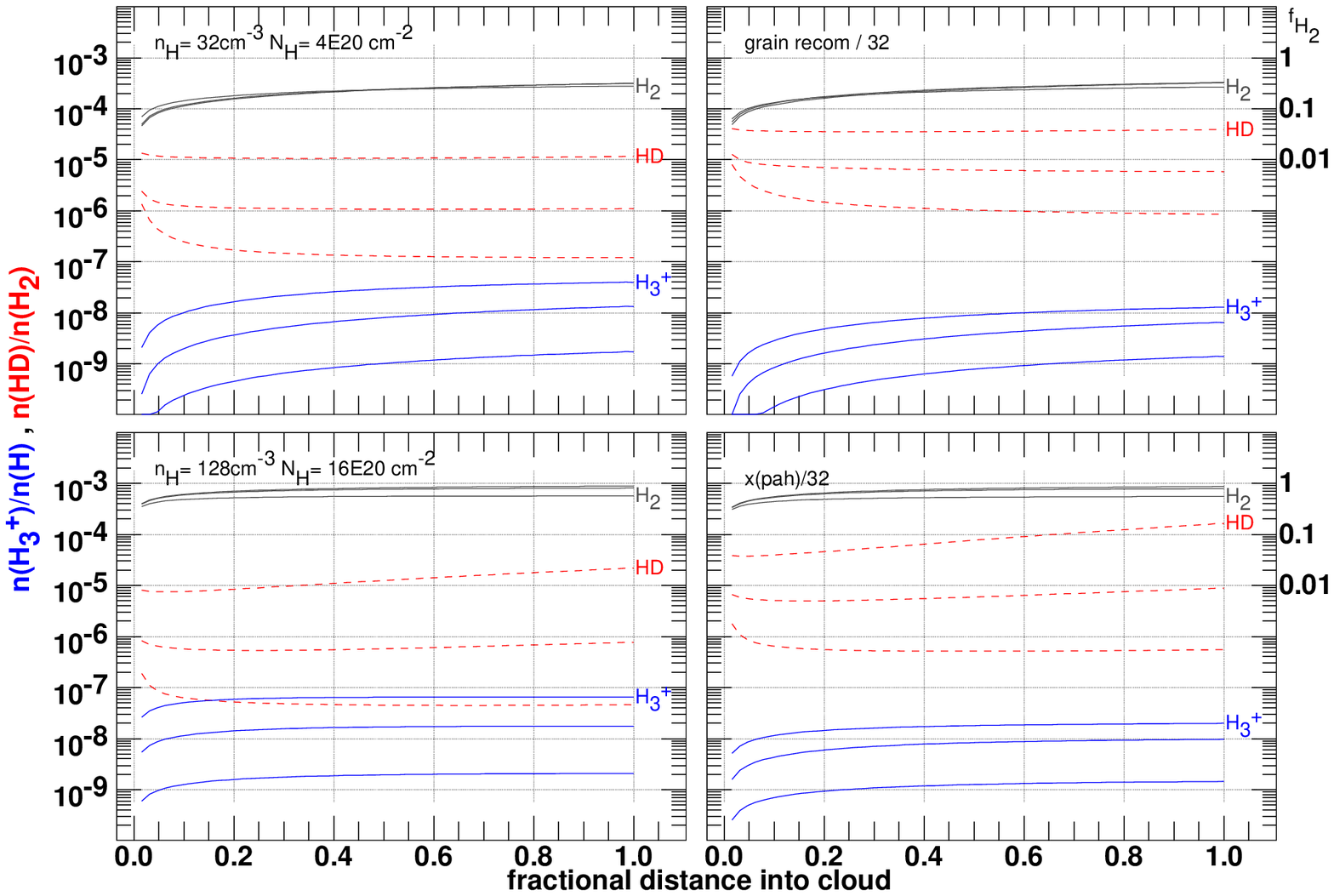,height=12cm}
\caption[]{Calculated HD, \HH\ and \H3p\ fractional abundances for three 
values of the cosmic ray ionization rate 
\zH $= 10^{-17}$\ps , $10^{-16}$\ps and $10^{-15}$\ps, 
for two uniformly illuminated spherical models. At top
is a `standard' Spitzer \citep{Spi68,Spi78} H I cloud
with n(H) $= 32~\pccc , {\rm N(H)} = 4 \times 10^{20}~\pcc$.
At bottom is a model having 4 times higher number and 
column density.  At right the effect of small grains
is diluted in each model, at top by simply scaling
down the grain neutralization rate, and at bottom by
scaling down the quantity of small grains (and so, all
associated effects).  The normalizations of the relative
abundances differ because N(\HH) is not known along most
lines of sight with known N(\H3p).
N(HD)/N(\HH) $\approx 1.2 \times 10^{-6}$ 
along unexceptional lines of sight (Fig. 4) and 
N(\H3p)/$(5.8\times10^{21}~\pcc \EBV) 
\approx 2.3\times10^{-8}$
in the survey of \cite{McCHin+02}.} 
\end{figure*}

\section{Implications and considerations for dense gas}

\subsection{Metal depletions and the problem of sulfur chemistry in dense gas}

The Solar elemental abundances of trace metals (Mg, Na, Fe, Si \ETC), 
$2-4\times10^{-5}$, while small compared to that of carbon, are much higher
than the implied ionization fractions in dense dark molecular gas (see just
below).  Thus, even a small ionization fraction in any one of such species 
could dominate the overall charge balance in dark clouds.

It was noted by \cite{OppDal74} that molecular ions will efficiently transfer 
charge to such metal atoms in dense gas. Because atomic ions recombine so much 
more slowly, in the gas (so the argument went), the overall ionization fraction 
is much increased in this way (if grain neutralization is neglected, as has 
been the case in this subject).  In turn, because rather low ionization 
fractions are required for maintenance of high \H3p abundances and efficient 
deuteration in dense gas, chemical modellers have typically adopted what is 
described as a 'low-metals' case, in which many metals are assumed to be 
strongly depleted from the gas phase ($\delta \approx 0.001$).  This includes 
sulfur (for example see \cite{NilHja+00}), although it, unlike the other 
species noted just above, does not undergo charge exchange with the most 
abundant molecular ions (\H3p, O\H3p\ and \hcop).

Such extreme sulfur depletion leads to several contradictions.  Unlike, say,
Mg, Na or Fe, sulfur is undepleted at lower extinction \citep{SavSem96} and 
the patterns of gas-phase depletion overall do not explain why it should be 
missing in denser gas.  More importantly, removing so much sulfur from the gas 
phase makes it impossible to form commonly-observed compounds like CS, SO, 
\ETC\ at the observed levels without harboring abundances of OH or 
O$_2$ which are 100 or more times what is observed, see \cite{Tur95} and 
\cite{NilHja+00}. The sulfur-bearing compounds are formed by reactions of 
free neutral sulfur atoms with these species and the abundance of the 
sulfur-bearing molecules scale linearly with the amount of free sulfur 
({\it ibid}).

Although some of the discussion of \cite{LepDal88} is mooted by recognition
that dissociative recombination of \H3p\ is efficient, and although they 
did not address sulfur explicitly, these authors demonstrated that grain 
neutralization renders the issue of metal depletion irrelevant to the free 
electron abundance in dense, dark molecular gas.  The charge carried by 
atomic ions is suppressed by grain neutralization; much of the entire
negative charge contribution resides on small grains (see Sect. 2.2 here), 
and the free electron density varies little (less than a factor two) whether 
metals are depleted or present.  In this case,  the abundance of gas phase 
sulfur may safely be assumed to be large enough that models with acceptably 
small OH and O$_2$ abundances -- when they emerge -- will easily produce 
enough of the sulfur-bearing molecules.  Additionally the gas phase metals
will be largely neutral, owing to grain neutralization.  This acts
to increase the abundances of most sulfur-bearing compounds because 
S\p\ is less chemically productive.  Consequences for the abundances of
molecules containing other metals largely remain to be explored.

\subsection{The atomic gas fraction in dense gas}

In a dark or molecular cloud, very nearly every cosmic ray ionization 
results in formation of an \H3p\ ion and the ultimate appearance of 
q = 2-3 hydrogen atoms \citep{CraDal78}.  Hydrogen atoms leave the gas 
phase by sticking to (and forming \HH\ on) large grains whose density 
is assumed proportional 
to n(H) = n(\HH)/2. The rate constant for this process is usually 
denoted by R and the equilibrium density of H-atoms in a dark, fully 
molecular gas is n(H I) $\approx$ q\zH/R, independent of n(\HH). Typically, 
R $\approx 3 \times 10^{-17}~\ccc$\ps\ is inferred in diffuse gas 
\citep{Jur74,Spi78,GryBou+02} {\ bf, so} n(H I) $ \approx 1 ~\pccc$, independent
of n(\HH), for the reference value of \zH.

It has been recognized since the work of \cite{AllRob76} that the 
fraction of atomic hydrogen inside dark clouds, while usually small, is 
still often much larger than expected for steady-state equilibrium between 
\HH\ formation at typical rates in cold gas and a cosmic ray ionization rate 
\zH\ $= 10^{-17}$\ps.  Observations of CO emission and H I self-absorption 
in low-latitude 21 cm profiles by \cite{LisBur+81,LisBur+84} suggested 
n(H I)/n(\HH) $\approx 0.005$ and similarly high values have recently been 
cited by \cite{WilLan+02} for some local dark clouds.  The latter authors 
invoked a process of turbulent diffusion to account for the H-fraction, 
while \cite{AllRob76} asserted that dark clouds were not old enough for 
full H I - \HH\ conversion (also see \cite{RawHar+02}).  

The higher values of \zH\ suggested by this work nominally imply an 
enhanced atomic hydrogen fraction in fully molecular gas, in agreement
with observation.  However, the 
rate of \HH\ formation is really well-known only in diffuse gas, where 
photoprocesses dominate and grains are incompletely dressed.  The rate 
of \HH\ formation in dark gas has only been interpreted in the context of the 
diffuse cloud data and/or relatively small \zH.  Theoretical models, which 
might account for changes in the formation rate when photodesorption is nil 
and  depletion proceeds, are largely lacking.

\subsection{The cosmic-ray heating rate in dense gas}

The standard calculation of heating in dense dark gas is that of 
\cite{Gol01} who used a (cosmic-ray) heating rate 
$10^{-27}$ n(\HH) erg $\pccc~\ps$.  Using the mean value of the heat
deposited  per primary event in a weakly ionized, fully-molecular gas 
calculated by \cite{CraDal78}, $\Delta$Q= 7 ev, we find that this heating 
rate corresponds to  \zH\ $= 4.5 \times 10^{-17}$\ps, some 4.5 times 
our reference value; \cite{Gol01}, without attribution, assumed 
$\Delta$Q= 20 ev.   
Thus, increasing our primary ionization rate by a factor of 20 would 
provide heating which is only a factor 4.5 greater than that in \cite{Gol01}.
The local cooling function varies as T$^{2.7\pm0.3}$ if gas-grain coupling 
is neglected and rather more rapidly if it is included.  A substantial 
increase in \zH\ (compared to our reference value) could be tolerated 
without inducing major changes in the temperature of dense gas; indeed, 
to heat molecular gas to the observed degree apparently requires a value 
of \zH\ which is at least 5 times larger than our reference value.

\section{Summary}

Gas-phase atomic ions are rapidly neutralized by the same population of small 
grains which heats the diffuse interstellar medium.  We considered the 
effects of such processes on various forms of hydrogen; protons in 
atomic gas; protons (deuterons), HD and \H3p\ in diffuse gas having
a molecular component, and H I in dense molecular clouds.

In atomic gas, limits on the cosmic-ray ionization rate \zH\ (the primary
rate per H-nucleus; see Table 1) from non-detection of 
low-frequency hydrogen recombination lines must be revised upward by such large 
factors that they are no longer restrictive.  If grain neutralization is
recognized, limits on \zH\ from the RRL are higher than the rate needed 
to ionize the intercloud medium in the absence of the soft X-ray flux, as in the
original theories of two-phase equilibrium.
Ion stage ratios like N(C II)/N(C I), N(Fe II)/N(Fe I) {\it etc.} 
provide accurate electron densities only when grain neutralization is fully 
taken into account.

HD and the newly-discovered \H3p\ may be shown to exist in the observed amounts, 
even in classical cool H I clouds of moderate density, if grain neutralization
proceeds and the cosmic-ray ionization rate has values 
\zH\ $\ga 2\times10^{-16}$\ps.  Models having lower values of \zH\ 
will reproduce HD if grain neutralization is neglected but fall grossly short
of explaining \H3p.

In denser gas, a higher cosmic ray ionization rate naturally accounts for the
(often) surprisingly high H I fractions, without invoking mechanisms such as
non-equilibrium \HH\ formation, small cloud ages, or turbulent diffusion
(all of which might still occur).  More importantly, we noted that inclusion of 
small-grain charging in dense gas removes the need for the 'low-metals' case of 
severe {\it ad hoc} sulfur depletion, thus allowing formation of sulfur-bearing 
molecules without grossly overproducing OH and O$_2$.

The present work represents no more than the barest first step at including
grain neutralization and its necessary consequences (like enhanced values of
\zH), into our understanding of the interstellar medium.

\begin{acknowledgements}

The National Radio Astronomy Observatory is operated by Associated Universites, Inc. 
under a cooperative agreement with the US National Science Foundation. Comments
of the referee, Dr. Mark Wolfire, are gratefully acknowledged.

\end{acknowledgements}


\begin{thebibliography}{69}
\expandafter\ifx\csname natexlab\endcsname\relax\def\natexlab#1{#1}\fi

\bibitem[{{Allen} \& {Robinson}(1976)}]{AllRob76}
{Allen}, M. \& {Robinson}, G.~W. 1976, ApJ, 207, 745

\bibitem[{{Amano}(1988)}]{Ama88}
{Amano}, T. 1988, ApJ, 329, L121

\bibitem[{{Bakes} \& {Tielens}(1994)}]{BakTie94}
{Bakes}, E. L.~O. \& {Tielens}, A. G. G.~M. 1994, ApJ, 427, 822

\bibitem[{{Barsuhn} \& {Walmsley}(1977)}]{BarWal77}
{Barsuhn}, J. \& {Walmsley}, C.~M. 1977, A\&A, 54, 345

\bibitem[{{Black} \& {Dalgarno}(1973)}]{BlaDal73}
{Black}, J.~H. \& {Dalgarno}, A. 1973, ApJ, 184, L101

\bibitem[{{Cravens} \& {Dalgarno}(1978)}]{CraDal78}
{Cravens}, T.~E. \& {Dalgarno}, A. 1978, ApJ, 219, 750

\bibitem[{{Crutcher}(1980)}]{Cru80}
{Crutcher}, R.~M. 1980, ApJ, 239, 549

\bibitem[{{D'Hendecourt} \& {Leger}(1987)}]{DHeLeg87}
{D'Hendecourt}, L.~B. \& {Leger}, A. 1987, A\&A, 180, L9

\bibitem[{{Doty} {et~al.}(2002){Doty}, {van Dishoeck}, {van der Tak}, \&
  {Boonman}}]{Dotvan+02}
{Doty}, S.~D., {van Dishoeck}, E.~F., {van der Tak}, F.~F.~S., \& {Boonman},
  A.~M.~S. 2002, A\&A, 389, 446

\bibitem[{{Draine} \& {Sutin}(1987)}]{DraSut87}
{Draine}, B.~T. \& {Sutin}, B. 1987, ApJ, 320, 803

\bibitem[{{Field} {et~al.}(1969){Field}, {Goldsmith}, \& {Habing}}]{FieGol+69}
{Field}, G.~B., {Goldsmith}, D.~W., \& {Habing}, H.~J. 1969, ApJ, 155, L149

\bibitem[{{Field} \& {Steigman}(1971)}]{FieSte71}
{Field}, G.~B. \& {Steigman}, G. 1971, ApJ, 166, 59+

\bibitem[{{Fitzpatrick}(1997)}]{Fit97}
{Fitzpatrick}, E.~L. 1997, ApJ, 482, L199

\bibitem[{{Fitzpatrick} \& {Spitzer}(1997)}]{FitSpi97}
{Fitzpatrick}, E.~L. \& {Spitzer}, L.~J. 1997, ApJ, 475, 623+

\bibitem[{{Geballe} {et~al.}(1999){Geballe}, {McCall}, {Hinkle}, \&
  {Oka}}]{GebMcC+99}
{Geballe}, T.~R., {McCall}, B.~J., {Hinkle}, K.~H., \& {Oka}, T. 1999, ApJ,
  510, 251

\bibitem[{{Goldsmith}(2001)}]{Gol01}
{Goldsmith}, P.~F. 2001, ApJ, 557, 736

\bibitem[{{Gry} {et~al.}(2002){Gry}, {Boulanger}, {Nehm{\' e}}, {Pineau des
  For{\^ e}ts}, {Habart}, \& {Falgarone}}]{GryBou+02}
{Gry}, C., {Boulanger}, F., {Nehm{\' e}}, C., {Pineau des For{\^ e}ts}, G.,
  {Habart}, E., \& {Falgarone}, E. 2002, A\&A, 391, 675

\bibitem[{{Jenkins} {et~al.}(1983){Jenkins}, {Jura}, \&
  {Loewenstein}}]{JenJur+83}
{Jenkins}, E.~B., {Jura}, M., \& {Loewenstein}, M. 1983, ApJ, 270, 88

\bibitem[{{Jura}(1974)}]{Jur74}
{Jura}, M. 1974, ApJ, 191, 375

\bibitem[{{Larsson} {et~al.}(1993){Larsson}, {Danared}, {Mowat}, {Sigray},
  {Sundstrom}, {Brostrom}, {Filevich}, {Kallberg}, {Mannervik}, \&
  {Rensfelt}}]{LarDan+93}
{Larsson}, M., {Danared}, H., {Mowat}, J.~R., {Sigray}, P., {Sundstrom}, G.,
  {Brostrom}, L., {Filevich}, A., {Kallberg}, A., {Mannervik}, S., \&
  {Rensfelt}, K.~G. 1993, Phys. Rev. Lett., 70, 430

\bibitem[{{Lee} {et~al.}(1996){Lee}, {Herbst}, {Pineau Des Forets}, {Roueff},
  \& {Le Bourlot}}]{LeeHer+96}
{Lee}, H.~H., {Herbst}, E., {Pineau Des Forets}, G., {Roueff}, E., \& {Le
  Bourlot}, J. 1996, A\&A, 311, 690

\bibitem[{{Lepp} \& {Dalgarno}(1988{\natexlab{a}})}]{LepDal88}
{Lepp}, S. \& {Dalgarno}, A. 1988{\natexlab{a}}, ApJ, 335, 769

\bibitem[{{Lepp} \& {Dalgarno}(1988{\natexlab{b}})}]{LepDal88a}
---. 1988{\natexlab{b}}, ApJ, 324, 553

\bibitem[{{Lepp} {et~al.}(1988){Lepp}, {Dalgarno}, {van Dishoeck}, \&
  {Black}}]{LepDal+88}
{Lepp}, S., {Dalgarno}, A., {van Dishoeck}, E.~F., \& {Black}, J.~H. 1988, ApJ,
  329, 418

\bibitem[{{Liszt}(2001)}]{Lis01}
{Liszt}, H. 2001, A\&A, 371, 698

\bibitem[{{Liszt}(2002)}]{Lis02}
---. 2002, A\&A, 389, 393

\bibitem[{{Liszt} \& {Lucas}(1999)}]{LisLuc99}
{Liszt}, H. \& {Lucas}, R. 1999, A\&A, 347, 258

\bibitem[{{Liszt}(1979)}]{Lis79}
{Liszt}, H.~S. 1979, ApJ, 233, L147

\bibitem[{{Liszt} {et~al.}(1981){Liszt}, {Burton}, \& {Bania}}]{LisBur+81}
{Liszt}, H.~S., {Burton}, W.~B., \& {Bania}, T.~M. 1981, ApJ, 246, 74

\bibitem[{{Liszt} {et~al.}(1984){Liszt}, {Burton}, \& {Xiang}}]{LisBur+84}
{Liszt}, H.~S., {Burton}, W.~B., \& {Xiang}, D.-L. 1984, A\&A, 140, 303

\bibitem[{{Liszt} \& {Lucas}(1994)}]{LisLuc94}
{Liszt}, H.~S. \& {Lucas}, R. 1994, ApJ, 431, L131

\bibitem[{{Liszt} \& {Lucas}(2000)}]{LisLuc00}
---. 2000, A\&A, 355, 333

\bibitem[{{Mathis} {et~al.}(1977){Mathis}, {Rumpl}, \& {Nordsieck}}]{MatRum+77}
{Mathis}, J.~S., {Rumpl}, W., \& {Nordsieck}, K.~H. 1977, ApJ, 217, 425

\bibitem[{{McCall} {et~al.}(2002){McCall}, {Hinkle}, {Geballe},
  {Moriarty-Schieven}, {Evans}, {Kawaguchi}, {Takano}, {Smith}, \&
  {Oka}}]{McCHin+02}
{McCall}, B.~J., {Hinkle}, K.~H., {Geballe}, T.~R., {Moriarty-Schieven}, G.~H.,
  {Evans}, N.~J., {Kawaguchi}, K., {Takano}, S., {Smith}, V.~V., \& {Oka}, T.
  2002, ApJ, 567, 391

\bibitem[{{Moos} {et~al.}(2002){Moos}, {Sembach}, {Vidal-Madjar}, {York},
  {Friedman}, {H{\' e}brard}, {Kruk}, {Lehner}, {Lemoine}, {Sonneborn}, {Wood},
  {Ake}, {Andr{\' e}}, {Blair}, {Chayer}, {Gry}, {Dupree}, {Ferlet}, {Feldman},
  {Green}, {Howk}, {Hutchings}, {Jenkins}, {Linsky}, {Murphy}, {Oegerle},
  {Oliveira}, {Roth}, {Sahnow}, {Savage}, {Shull}, {Tripp}, {Weiler}, {Welsh},
  {Wilkinson}, \& {Woodgate}}]{MooCow02}
{Moos}, H.~W., {Sembach}, K.~R., {Vidal-Madjar}, A., {York}, D.~G., {Friedman},
  S.~D., {H{\' e}brard}, G., {Kruk}, J.~W., {Lehner}, N., {Lemoine}, M.,
  {Sonneborn}, G., {Wood}, B.~E., {Ake}, T.~B., {Andr{\' e}}, M., {Blair},
  W.~P., {Chayer}, P., {Gry}, C., {Dupree}, A.~K., {Ferlet}, R., {Feldman},
  P.~D., {Green}, J.~C., {Howk}, J.~C., {Hutchings}, J.~B., {Jenkins}, E.~B.,
  {Linsky}, J.~L., {Murphy}, E.~M., {Oegerle}, W.~R., {Oliveira}, C., {Roth},
  K., {Sahnow}, D.~J., {Savage}, B.~D., {Shull}, J.~M., {Tripp}, T.~M.,
  {Weiler}, E.~J., {Welsh}, B.~Y., {Wilkinson}, E., \& {Woodgate}, B.~E. 2002,
  Astrophys. J., Suppl. Ser., 140, 3

\bibitem[{{Morton}(1975)}]{Mor75}
{Morton}, D.~C. 1975, ApJ, 197, 85

\bibitem[{{Nillson} {et~al.}(2000){Nillson}, {Hjalmarson}, {Bergman}, \&
  {Millar}}]{NilHja+00}
{Nillson}, A., {Hjalmarson}, P., {Bergman}, P., \& {Millar}, T. 2000, A\&A,
  358, 257

\bibitem[{{O'Donnell} \& {Watson}(1974)}]{ODoWat74}
{O'Donnell}, E. \& {Watson}, W.~D. 1974, ApJ, 191, 89

\bibitem[{{Omont}(1986)}]{Omo86}
{Omont}, A. 1986, A\&A, 164, 159

\bibitem[{{Oppenheimer} \& {Dalgarno}(1974)}]{OppDal74}
{Oppenheimer}, M. \& {Dalgarno}, A. 1974, ApJ, 192, 29

\bibitem[{{Payne} {et~al.}(1994){Payne}, {Anantharamaiah}, \&
  {Erickson}}]{PayAna+94}
{Payne}, H.~E., {Anantharamaiah}, K.~R., \& {Erickson}, W.~C. 1994, ApJ, 430,
  690

\bibitem[{{Payne} {et~al.}(1984){Payne}, {Salpeter}, \& {Terzian}}]{PaySal+84}
{Payne}, H.~E., {Salpeter}, E.~E., \& {Terzian}, Y. 1984, Astron. J., 89, 668

\bibitem[{{Rawlings} {et~al.}(2002){Rawlings}, {Hartquist}, {Williams}, \&
  {Falle}}]{RawHar+02}
{Rawlings}, J.~M.~C., {Hartquist}, T.~W., {Williams}, D.~A., \& {Falle},
  S.~A.~E.~G. 2002, A\&A, 391, 681

\bibitem[{{Savage} {et~al.}(1992){Savage}, {Cardelli}, \& {Sofia}}]{SavCar+92}
{Savage}, B.~D., {Cardelli}, J.~A., \& {Sofia}, U.~J. 1992, ApJ, 401, 706

\bibitem[{{Savage} {et~al.}(1977){Savage}, {Drake}, {Budich}, \&
  {Bohlin}}]{SavDra+77}
{Savage}, B.~D., {Drake}, J.~F., {Budich}, W., \& {Bohlin}, R.~C. 1977, ApJ,
  216, 291

\bibitem[{{Savage} \& {Sembach}(1996)}]{SavSem96}
{Savage}, B.~D. \& {Sembach}, K.~R. 1996, Ann. Rev. Astrophys. Astron., 34, 279

\bibitem[{{Shaver}(1976)}]{Sha76}
{Shaver}, P.~A. 1976, Astrophysics, 49, 149

\bibitem[{{Snow}(1975)}]{Sno75}
{Snow}, T.~P. 1975, ApJ, 201, L21

\bibitem[{{Snow}(1977)}]{Sno77}
---. 1977, Astrophysics, 216, 724

\bibitem[{{Sonnentrucker} {et~al.}(2002){Sonnentrucker}, {Friedman}, {Welty},
  {York}, \& {Snow}}]{SonFri+02}
{Sonnentrucker}, P., {Friedman}, S.~D., {Welty}, D.~E., {York}, D.~G., \&
  {Snow}, T.~P. 2002, ApJ, 576, 241

\bibitem[{{Spitzer}(1968)}]{Spi68}
{Spitzer}, L. 1968, {Diffuse matter in space} (New York: Interscience
  Publication, 1968)

\bibitem[{{Spitzer}(1978)}]{Spi78}
---. 1978, Physical processes in the interstellar medium (New York
  Wiley-Interscience, 1978. 333 p.)

\bibitem[{{Spitzer} {et~al.}(1974){Spitzer}, {Cochran}, \&
  {Hirshfeld}}]{SpiCoc+74}
{Spitzer}, L., {Cochran}, W.~D., \& {Hirshfeld}, A. 1974, Astrophys. Space.
  Sci., 28, 373

\bibitem[{{Stancil} {et~al.}(1998){Stancil}, {Lepp}, \& {Dalgarno}}]{StaLep+98}
{Stancil}, P.~C., {Lepp}, S., \& {Dalgarno}, A. 1998, Astrophysics, 509, 1

\bibitem[{{Stancil} {et~al.}(1999){Stancil}, {Schultz}, {Kimura}, {Gu},
  {Hirsch}, \& {Buenker}}]{StaSch+99}
{Stancil}, P.~C., {Schultz}, D.~R., {Kimura}, M., {Gu}, J.-P., {Hirsch}, G., \&
  {Buenker}, R.~J. 1999, Astron. Astrophys. Suppl. Ser., 140, 225

\bibitem[{{Sundstrom} {et~al.}(1994){Sundstrom}, {Mowat}, {Danared}, {Sigray},
  {Brostrom}, {Filevich}, {Kallberg}, {Mannervik}, {Rensfelt}, \&
  {Larsson}}]{SunMow+94}
{Sundstrom}, G., {Mowat}, J.~R., {Danared}, H., {Sigray}, P., {Brostrom}, L.,
  {Filevich}, A., {Kallberg}, A., {Mannervik}, S., {Rensfelt}, K.~G., \&
  {Larsson}, M. 1994, Science, 263, 785

\bibitem[{{Taylor} \& {Cordes}(1993)}]{TayCor93}
{Taylor}, J.~H. \& {Cordes}, J.~M. 1993, ApJ, 411, 674

\bibitem[{{Turner}(1995)}]{Tur95}
{Turner}, B.~E. 1995, ApJ, 455, 556

\bibitem[{{van der Tak} \& {van Dishoeck}(2000)}]{vanvan+00}
{van der Tak}, F.~F.~S. \& {van Dishoeck}, E.~F. 2000, A\&A, 358, L79

\bibitem[{{Watson}(1973)}]{Wat73}
{Watson}, W.~D. 1973, ApJ, 182, L73

\bibitem[{{Weingartner} \& {Draine}(2001{\natexlab{a}})}]{WeiDra01}
{Weingartner}, J.~C. \& {Draine}, B.~T. 2001{\natexlab{a}}, ApJ, 548, 392

\bibitem[{{Weingartner} \& {Draine}(2001{\natexlab{b}})}]{WeiDra01a}
---. 2001{\natexlab{b}}, ApJ, 563, 842

\bibitem[{{Weingartner} \& {Draine}(2001{\natexlab{c}})}]{WeiDra01Supp}
---. 2001{\natexlab{c}}, Astrophys. J., Suppl. Ser., 134, 263

\bibitem[{{Welty} \& {Hobbs}(2001)}]{WelHob01}
{Welty}, D.~E. \& {Hobbs}, L.~M. 2001, Astrophys. J., Suppl. Ser., 133, 345

\bibitem[{{Willacy} {et~al.}(2002){Willacy}, {Langer}, \& {Allen}}]{WilLan+02}
{Willacy}, K., {Langer}, W.~D., \& {Allen}, M. 2002, ApJ, 573, L119

\bibitem[{{Wilson} {et~al.}(1993){Wilson}, {Mauersberger}, {Muders},
  {Przewodnik}, \& {Olano}}]{WilMau+93}
{Wilson}, T.~L., {Mauersberger}, R., {Muders}, D., {Przewodnik}, A., \&
  {Olano}, C.~A. 1993, A\&A, 280, 221

\bibitem[{{Wolfire} {et~al.}(2003){Wolfire}, {McKee}, {Hollenbach}, \&
  {Tielens}}]{WolMcK+03}
{Wolfire}, M., {McKee}, C.~F., {Hollenbach}, D., \& {Tielens}, A.~G.~G.~M.
  2003, Ap. J. {\it in press}

\bibitem[{{Wolfire} {et~al.}(1995){Wolfire}, {Hollenbach}, {McKee}, {Tielens},
  \& {Bakes}}]{WolHol+95}
{Wolfire}, M.~G., {Hollenbach}, D., {McKee}, C.~F., {Tielens}, A. G. G.~M., \&
  {Bakes}, E. L.~O. 1995, ApJ, 443, 152

\bibitem[{{Wright} \& {Morton}(1979)}]{WriMor79}
{Wright}, E.~L. \& {Morton}, D.~C. 1979, ApJ, 227, 483

\end{thebibliography}
\end{document}